\documentclass[11pt,twoside]{article}
\usepackage{asp2004,graphicx,lscape}
\def\Ni{$^{56}$Ni}
\def\Co{$^{56}$Co}
\begin{document}
\title{Consistent Radiative Transfer Models
including Time Dependent Energy Deposition for Type Ia Supernovae}
\author{P. Hultzsch, D. Sauer, A.W\kern-.2em.\,A. Pauldrach, T. Hoffmann}
\affil{MPI for Astrophysics, Karl-Schwarzschild-Str.~1, D-85741
Garching
\\
University Observatory Munich, Scheinerstr.~1, D-81679 Munich}

\begin{abstract}
Many aspects of the explosion mechanism of Type Ia supernovae (SN
Ia) still remain unclear -- causing uncertainties in the derived
cosmological parameters. Realistic models of the generation and
transport of radiation in the ejecta are required which link
theoretical explosion models to observations. We aim to construct
theoretical spectra and light curves from consistent radiative
transfer models that allow to study the dependence of observable
features on the physical parameters of the explosion.
\end{abstract}

\subsection*{1.~~~Introduction}
To minimize the systematic errors of cosmological quantities deduced
from SN~Ia it is essential to have a detailed understanding of the
physics involved in these explosions. An important tool are
realistic models of the radiation transport within the ejected
material which represent the crucial link between theoretical
explosion models and observations. The physical conditions within
the expanding SN~Ia ejecta, however, make this extremely difficult
and are the reason that models which allow a reliable, quantitative
analysis are still missing. High radiation densities acting on an
environment of low matter density, as well as the energy input by
$\gamma$-rays and positrons within the atmosphere, do not allow for
the simplifying assumptions of LTE. Thus a consistent model requires
at minimum the solution of the full non-LTE problem
\citep[cf.][]{pauldrach96}.  Here we outline our project to
construct a more consistent theoretical description of SN~Ia
radiative transfer and describe first steps.

\subsection*{2.~~~The Model}
\begin{figure}[h!]
\begin{minipage}{.62\textwidth}
\includegraphics[width=\textwidth]{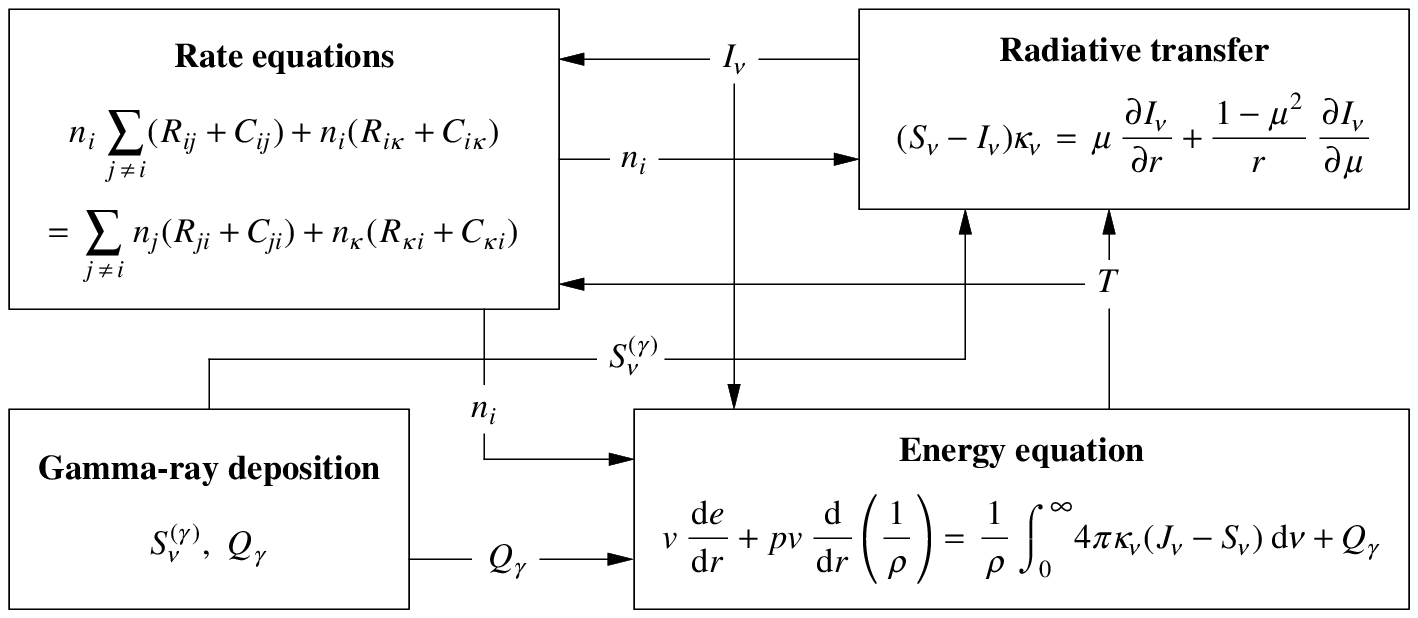}
\end{minipage}
\hfill
\begin{minipage}{.36\textwidth}
\includegraphics[angle=90,width=\textwidth]{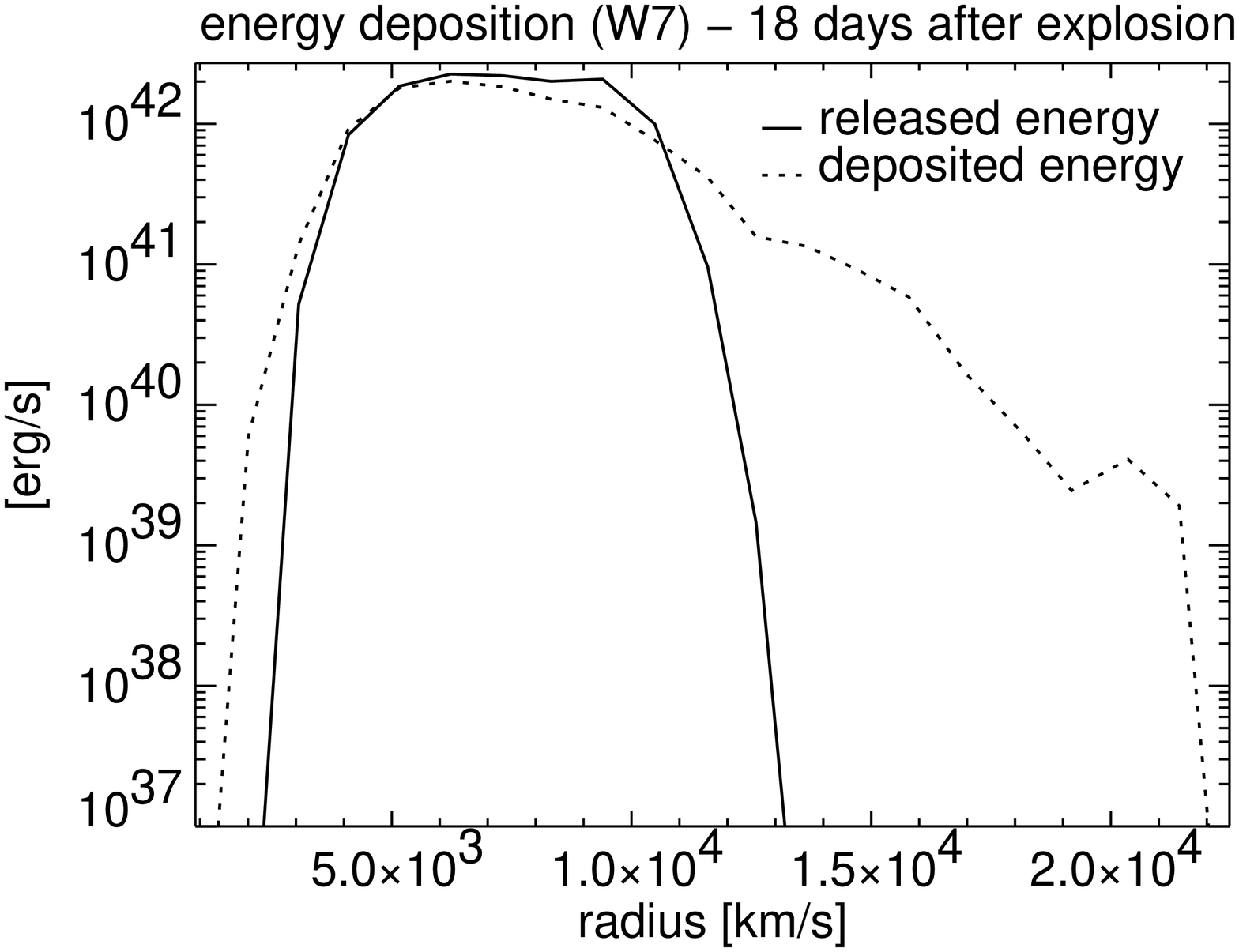}
\end{minipage}
\caption{Sketch of the basic physics treated in the NLTE model
(left) and a snapshot of the energy deposition rate as function of
radius (right).} \label{fig12}
\end{figure}

In our approach the description of radiative processes in supernova
ejecta is split into two problems: First, the time dependent
description of the deposition of radiative energy from the $\gamma$
photons from the {\Ni} and {\Co} decay as well as the trapping of
photons in highly opaque, expanding material that leads to the
characteristic shape of the light curve. Second, the actual
formation of the spectrum that needs to consider only the outer
parts of the ejecta where radiation is actually able to emerge
(figure~\ref{fig12}). Here the time scales for interaction of
photons with matter become small compared to the expansion time
scale so this part can be treated in a time independent model.  In
both parts we assume spherical symmetry of the object.

{\it Time dependent energy deposition.} In our approach the time
dependent part treated with a Monte Carlo (MC) light curve code
\citep{cappellaro97} that simulates the propagation and deposition
of the $\gamma$ photons and positrons within the ejecta using an
approximate, wavelength independent opacity. The resulting energy
deposition in the ejecta is distributed to optical photons. The
random walk of these photons on their way through the ejecta is
followed until they reach low opacity regions from where they can
escape. The amount of emerging energy per time interval determines
the light curve of the SN and sets the luminosity  for the synthetic
spectrum at a given epoch.

{\it NLTE models and spectral synthesis.} The energy deposition rate
at a certain epoch is used as an input to the stationary but much
more detailed computation of the radiation transport in full NLTE.
This model treats the outer part of the ejecta, assuming a
``photosphere'' as lower boundary. The code used here was originally
developed for the analysis of spectra of hot stars with radiation
driven winds \citep{pauldrach01} and is being modified in order to
treat the physical conditions in SN~Ia. The code provides a
consistent solution of the NLTE rate equations for all relevant
elements. The radiation transfer is calculated in detail taking into
account all significant sources of opacity and emission. Also, a
proper treatment of line blocking and blanketing effects is
included.  The radiation field at the outer edge of the fully
converged model represents the synthetic spectrum of the object, and
can be directly compared to observations.

\subsection*{3.~~~Status and first results}
So far we have derived simplified models, assuming all energy
deposition occurs below the photosphere, to test the code for
physical conditions in SN~Ia. Result of this analysis was that even
in deep layers the approximation of LTE does not give an appropriate
description of the radiation field \citep{sauer03}. This causes the
commonly used diffusion approximation as a lower boundary for the
radiative transport equations to break down, resulting in
unrealistic temperatures and a too large flux in the red part of the
spectrum. Adjustments to this boundary condition have been
investigated and are currently being tested. The next step will be
to incorporate the results computed with the MC code in the NLTE
calculations.


\begin{thebibliography}{}
\bibitem[Cappellaro et al.(1997)]{cappellaro97}
{Cappellaro}, E.~et al.~(1997), A\&A, 328, 203
\bibitem[Pauldrach et al.(1996)]{pauldrach96}
{Pauldrach}, A.W.A. et al.~(1996), A\&A, 312, 525
\bibitem[Pauldrach et al.(2001)]{pauldrach01}
{Pauldrach}, A.W.A., {Hoffmann}, T.L., and {Lennon}, M.~(2001),
A\&A, 375, 161
\bibitem[Sauer et al.(2003)]{sauer03}
{Sauer}, D.~et al.~(2003), Proc.~IAU Colloq.~192, astro-ph/0410703
\end{thebibliography}
\end{document}